\documentclass[conference]{IEEEtran}
\IEEEoverridecommandlockouts
\usepackage{hyperref}
\usepackage{cite}
\usepackage{amsmath,amssymb,amsfonts}
\usepackage{algorithmic}
\usepackage{graphicx}
\usepackage{textcomp}
\usepackage{xcolor}
\usepackage[caption=false,font=footnotesize]{subfig}
\usepackage{booktabs}
\usepackage{enumitem}
\setlist[itemize]{leftmargin=*}
\def\BibTeX{{\rm B\kern-.05em{\sc i\kern-.025em b}\kern-.08em
    T\kern-.1667em\lower.7ex\hbox{E}\kern-.125emX}}
\newcommand{\dist}{\mathcal{D}}
\newcommand{\distm}[1]{\dist^{(#1)}}
\newcommand{\eleproc}[2][p]{\mathcal{I}^{(#1)}\left(#2\right)}
\newcommand{\eleprocd}[1][p]{\mathcal{I}^{(#1)}}
\newcommand{\eleprocm}[3][p]{\mathcal{I}^{(#1)}_{#2}\left(#3\right)}
\newcommand{\eleprocmd}[2][p]{\mathcal{I}^{(#1)}_{#2}}
\newcommand{\numelem}[3][p]{I^{(#1)}_{#2}\left(#3\right)}
\newcommand{\numelemd}[2][p]{I^{(#1)}_{#2}}
\newcommand{\procsp}[2][p]{\mathcal{P}^{(#1)}\left(#2\right)}

\newcommand{\sizem}[1]{I_{#1}}
\newcommand{\Cost}{\mathrm{Cost}}
\newcommand{\Costl}[2]{\Cost_{#1}(#2)}
\newcommand{\Shuffle}[2]{\mathrm{Shuffle}(#1, #2)}

\newcommand{\MeshI}{\mathit{1}K}
\newcommand{\MeshII}{\mathit{2}K}

\begin{document}

\title{Improving Strong-Scaling of CNN Training by Exploiting Finer-Grained Parallelism}

\author{\IEEEauthorblockN{Nikoli Dryden\IEEEauthorrefmark{1}\IEEEauthorrefmark{2}\thanks{The first and second authors contributed equally.}, Naoya Maruyama\IEEEauthorrefmark{1}, Tom Benson\IEEEauthorrefmark{1}, Tim Moon\IEEEauthorrefmark{1}, Marc Snir\IEEEauthorrefmark{2}, Brian Van Essen\IEEEauthorrefmark{1}}
\IEEEauthorblockA{\IEEEauthorrefmark{1}Lawrence Livermore National Laboratory\\
\{maruyama3,benson31,moon13,vanessen1\}@llnl.gov}
\IEEEauthorblockA{\IEEEauthorrefmark{2}Department of Computer Science\\
University of Illinois at Urbana-Champaign\\
\{dryden2,snir\}@illinois.edu}}

\maketitle

\begin{abstract}
Scaling CNN training is necessary to keep up with growing datasets and reduce training time. We also see an emerging need to handle datasets with very large samples, where memory requirements for training are large. Existing training frameworks use a data-parallel approach that partitions samples within a mini-batch, but limits to scaling the mini-batch size and memory consumption makes this untenable for large samples. We describe and implement new approaches to convolution, which parallelize using spatial decomposition or a combination of sample and spatial decomposition. This introduces many performance knobs for a network, so we develop a performance model for CNNs and present a method for using it to automatically determine efficient parallelization strategies.

We evaluate our algorithms with microbenchmarks and image classification with ResNet-50. Our algorithms allow us to prototype a model for a mesh-tangling dataset, where sample sizes are very large. We show that our parallelization achieves excellent strong and weak scaling and enables training for previously unreachable datasets.
\end{abstract}

\IEEEpeerreviewmaketitle


\section{Introduction}
\label{sec:intro}
A key factor in the success of deep learning~\cite{lecun2015deep} has been the availability of sufficient compute power to meet the demands of training deep networks. GPUs have been particularly important in enabling this~\cite{raina2009large,krizhevsky2012imagenet}. Training a modern convolutional neural network (CNN) to convergence typically involves many iterations through large datasets, which can take days to weeks. Clusters of GPUs are often employed to help train models in more reasonable timeframes. This is especially important for researchers developing novel models or working with novel datasets, where rapid turnaround is needed. Further, dataset sizes continue to grow~\cite{sun2017revisiting,mahajan2018exploring}, deep learning is being applied to new domains such as medicine and physics, and models are becoming more complex~\cite{canziani2016analysis}. This necessitates continued scaling and performance improvements.

GPUs are becoming increasingly more prevalent in HPC systems; more than half of the new FLOPS added in the June, 2018 Top-500 list were due to GPUs~\cite{top500june2018}. These systems are well-suited to training CNNs, and there has been much interest in this. Commercial firms are also increasingly turning to large clusters for training. It is therefore important to scale training---for datasets with samples of all sizes---to such systems, allowing researchers to take advantage of their significant compute power.

Classic datasets have relatively small samples. ImageNet~\cite{ILSVRC15} images, for example, are typically resized to be $224 \times 224$ with three color channels. Emerging datasets may have significantly larger samples, especially those from large-scale numerical simulations. We consider here a dataset for detecting mesh-tangling in numerical simulations, where each sample is a $2048 \times 2048$ image with 18 channels, generated from a hydrodynamics code. Each sample requires ${\sim}288$ MiB (in single-precision). Larger sample sizes are easily possible depending on the simulation size and resolution, including the case where a \emph{single} sample is too large to fit in GPU memory. This is typical of the scale of data produced by HPC applications. There are other domains where samples may be large, including medical imagery~\cite{litjens2017survey}, drug discovery~\cite{chen2018rise}, video~\cite{oh2011large}, and satellite imagery~\cite{mundhenk2016large}. As training requires the activations of each layer to be preserved until backpropagation completes, and the size of activations depends on the input data size, large samples result in very high pressure on GPU memory (e.g. 16 or 32 GB on a V100).

Current deep learning software typically employs a data-parallel approach to scaling, wherein the samples in a mini-batch are partitioned between processors. Typical mini-batch sizes are 64-1000 samples, with 256 common. Processors then perform forward and backpropagation independently, only needing to synchronize parameter updates. However, data-parallel scaling is limited by the number of samples in a mini-batch, which can be difficult to increase due to generalization and convergence issues~\cite{keskar2017large,shallue2018measuring}. Recent work has been successful at training on ImageNet with large mini-batches (e.g. 8192+ samples)~\cite{goyal2017accurate,you2018imagenet}, but it is unclear if these techniques generalize to other datasets. Further, data-parallel scaling cannot reduce memory usage beyond what is required for a single sample, and therefore is not viable for very large samples.

We address both scaling and memory issues by exploiting parallelism in convolutional layers beyond data-parallelism (which we refer to as \emph{sample} parallelism). Conceptually, we think of a convolutional layer as being specified by five dimensions: samples, height, width, channels, and filters. (See Section~\ref{subsec:background-cnns}. Generalizing to additional spatial dimensions is not hard, and we ignore the convolutional kernel size as it is typically small in modern CNNs.) Sample parallelism partitions layers only along the first dimension; we describe and implement algorithms to additionally partition the spatial dimensions by splitting up input samples. This enables additional parallelism, but requires extra communication in forward and backpropagation, which can make them less advantageous in some situations compared to pure sample parallelism. We also discuss algorithms for partitioning the channel and filter dimensions. These, and our implementation in LBANN~\cite{van2015lbann}, are described in detail in Sections~\ref{sec:algos} and \ref{sec:implementation}.

These additional parallelism techniques make it complicated to determine how to parallelize a network to achieve good speedup on a system. Section~\ref{sec:perfmodel} introduces a technique to automatically find good parallel execution strategies. Given a platform and a CNN architecture, our system uses a performance model to determine promising ways to parallelize the network, accounting for memory requirements.

In the remainder of the paper, we evaluate our algorithms with ResNet-50~\cite{he2016deep} on ImageNet-1K classification~\cite{ILSVRC15} and proof-of-concept models for our mesh-tangling dataset. Training models on this dataset was previously infeasible due to memory requirements, so this is an initial demonstration of capability rather than a final model. Future models and evaluations will undoubtably improve upon this now that training is feasible.

We summarize our contributions as follows:
\begin{itemize}
\item We describe and implement algorithms for parallelizing convolutional layers by exploiting parallelism available from sample and spatial decompositions. We additionally describe extensions to channel and filter decompositions.
\item We present a performance model for CNNs and an approach for determining good parallel execution strategies for CNNs.
\item We comprehensively evaluate these implementations with microbenchmarks and end-to-end training.
\item We demonstrate the feasibility of training on a dataset with very large samples.
\end{itemize}

\section{Background and notation}
\label{sec:background}
Here we provide a brief overview of relevant background on CNNs and performance modeling, and introduce our notation. We assume the reader has a basic familiarity with deep learning.

\subsection{Convolutional neural networks}
\label{subsec:background-cnns}
Consider a 2D convolutional layer, with $F$ filters of size $K \times K$, stride $S$, and padding $P$. For simplicity, we assume $K$ is odd, and write $O = \lfloor K / 2 \rfloor$ to be the number of filter entries on either side of the center. Its input consists of $N$ samples, each with $C$ channels, height $H$, and width $W$. Each filter is applied to each sample, resulting in $N$ outputs with $F$ channels, height $\tilde{H}$, and width $\tilde{W}$. These quantities define the layer, which we think of as six 4D tensors: a $N \times C \times H \times W$ input $x$, a $F \times C \times K \times K$ weights $w$, and a $N \times F \times \tilde{H} \times \tilde{W}$ output $y$, plus the associated gradients. We do not consider alternate storage layouts (e.g. $NHWC$) in this paper. The CNN is trained with a loss function $L$.

To simplify notation, we assume $S = 1$ and ``same'' padding. Subscripts of $x$ and $dL/dy$ may be ``out of range'' (e.g. negative); we assume these are handled with padding (these assumptions are not necessary for our work). Forward propagation is given by
\begin{equation}
y_{k,f,i,j} = \sum_{c=0}^{C-1} \sum_{a=-O}^O \sum_{b=-O}^O x_{k,c,i+a,j+b} w_{f,c,a+O,b+O} \label{eq:fp}
\end{equation}

In backpropagation, a layer has input $dL/dy$ (of shape $N \times F \times \tilde{H} \times \tilde{W}$), which we refer to as an \emph{error signal}. It computes the gradients for its weights, $dL/dw$, and an error signal for the next layer, $dL/dx$ as
\begin{equation}
\frac{dL}{dw_{f,c,a,b}} = \sum_{k=0}^{N-1} \sum_{i=0}^{H-1} \sum_{j=0}^{W-1} \frac{dL}{dy_{k,f,i,j}} x_{k,c,i+a-O,j+b-O} \label{eq:bp-weights}
\end{equation}
\begin{equation}
\frac{dL}{dx_{k,c,i,j}} = \sum_{f=0}^{F-1} \sum_{a=-O}^O \sum_{b=-O}^O \frac{dL}{dy_{k,f,i-a,j-b}} w_{f,c,a+O,b+O}. \label{eq:bp-data}
\end{equation}

Note that the common data-parallel (sample-parallel) formulation for convolutional layers can be immediately seen from these formulas: the $N$ dimension is partitioned, resulting in local computations for $y$ and $dL/dx$, and the summation in $dL/dw$ is aggregated with an allreduce.

We do not focus on the implementation of convolution itself and instead rely on cuDNN~\cite{chetlur2014cudnn} for an optimized implementation.

Other layers, such as batch normalization~\cite{ioffe2015batch}, ReLUs, pooling, and fully-connected (FC) layers are also present in CNNs. We do not focus on their details, as modifications are either simple or similar to those for convolutional layers. For FC layers, we use a model-parallel formulation based on distributed matrix products~\cite{van2015lbann}.

\subsection{Performance modeling}
\label{subsec:background-perfmodel}
We make use of analytic models for some of our performance estimates, particularly communication. We use a linear model~\cite{fraigniaud1994methods} for communication, where $\alpha$ is the latency and $\beta$ is the inverse bandwidth. Then the cost to send a message between two nodes is $\alpha + \beta n$. We additionally assume that the network is full-duplex and that there is no interference.

Collective communication operations such as allreduce will be important for some operations; for these, we use the performance models of Thakur et al~\cite{thakur2005optimization}. For distributed matrix multiplication, we use the  performance models developed for the Elemental library~\cite{schatz2016parallel}.

\subsection{Notation}
\label{subsec:background-notation}
We now define some notation for distributed tensors that will be used throughout this paper. Our notation is heavily based on the tensor notation developed for the FLAME project~\cite{poulson2013elemental,schatz2015distributed,schatz2016parallel}.

A tensor is an $M$-dimensional array, where the size of dimension $m$ is $\sizem{m}$, and we write $I = (\sizem{0}, \ldots, \sizem{M-1})$ to refer to the shape of an entire tensor.

\subsubsection{Distributions}
Let $P$ be the set of processors over which we distribute data. In general, we assume that every processor has the same amount of data (excepting minor imbalances due to divisibility) for load-balancing reasons, but this is not required.

We refer to the distribution of a tensor as $\dist$, and of a particular dimension $m$ as $\distm{m}$. A distribution $\distm{m}$ can be thought of as an assignment of indices in that dimension to processors. Then a distribution for a tensor is $\dist = \left(\distm{0}, \ldots, \distm{M-1}\right)$. We require that every element of a tensor be assigned to at least one processor.

\subsubsection{Processor and index sets}
It is helpful to be able to refer to different sets of processors and indices for a distribution. We use $\eleproc{\dist}$ to refer to the set of elements on processor $p$ under distribution $\dist$, and $\eleprocm{m}{\distm{m}}$ to the indices in dimension $m$ on $p$. The number of indices in $m$ on $p$ is $\numelem{m}{\distm{m}}$. We omit the distribution when it is clear from context. We write $\procsp{\distm{m_0}, \ldots}$ for the set of processors that have the same indices as $p$ under the distributions.


\subsubsection{CNNs}
We think of a CNN as a directed acyclic graph, where a layer may have multiple parents or children (e.g. residual connections). Each layer $\ell_i$ has six tensors associated with it: input $x_i$, output $y_i$, weights $w_i$, input error signal $dL/dy_i$, weight gradients $dL/dw_i$, and output error signal $dL/dx_i$. (If a layer has no parameters, $w_i$ and $dL/dw_i$ are empty.) For a ``line'' network architecture, the output from the previous layer is $x_i = y_{i-1}$ and the error signal from the subsequent layer is $dL/dy_i = dL/dx_{i+1}$.

Many of the dimensions for a layer's tensors are the ``same''; for example, the sample dimension $N$ of a convolutional layer's inputs and outputs. To avoid communication, it makes sense for dimensions to be distributed the same way in both tensors. Thus we will overload our notation to allow $\dist$ to refer to the distribution of a layer, with the distributions for each dimension applied to the appropriate tensors.


\section{Distributed memory convolution algorithms}
\label{sec:algos}
We now present our algorithms for distributed-memory convolution. We refer to exploiting parallelism along a particular dimension as ``parallelizing'' that dimension. For convolutional layers, this results in five approaches: sample, height and width (together, spatial), channel, and filter parallelism. Sample parallelism is simply the familiar data-parallel approach. Note that these approaches are not mutually exclusive.

We begin by describing spatial parallelism, together with sample parallelism, and then discuss our implementation. Our algorithms exactly replicate convolution as if it were performed on a single GPU (up to floating point accumulation issues). We also briefly discuss channel and filter parallelism, but we defer implementation to future work.

One assumption we require is that spatial dimensions are distributed in a blocked manner. This is because applying convolution at a point requires spatially adjacent data, and a non-blocked distribution would require extensive communication.

\subsection{Sample and spatial parallelism}
\label{subsec:algo-spatial}

\begin{figure*}
  \centering
  \begin{minipage}{0.6\textwidth}\centering
    \subfloat[Sample parallelism\label{fig:algo-sample}]
    {\includegraphics[scale=0.4]{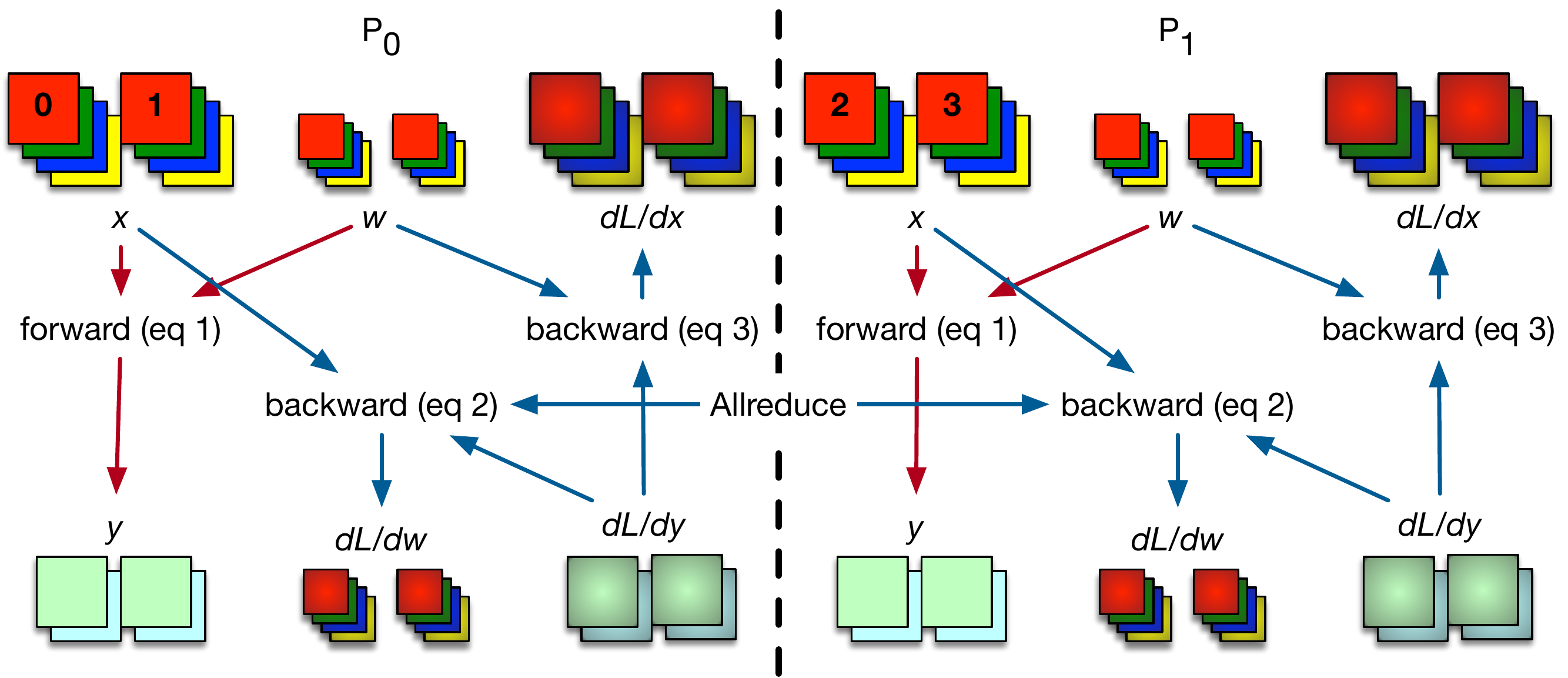}}
  \end{minipage}
  \begin{minipage}{0.3\textwidth}\centering
    \subfloat[Spatial parallelism halo exchange\label{fig:algo-spatial}]
    {\includegraphics[scale=0.35]{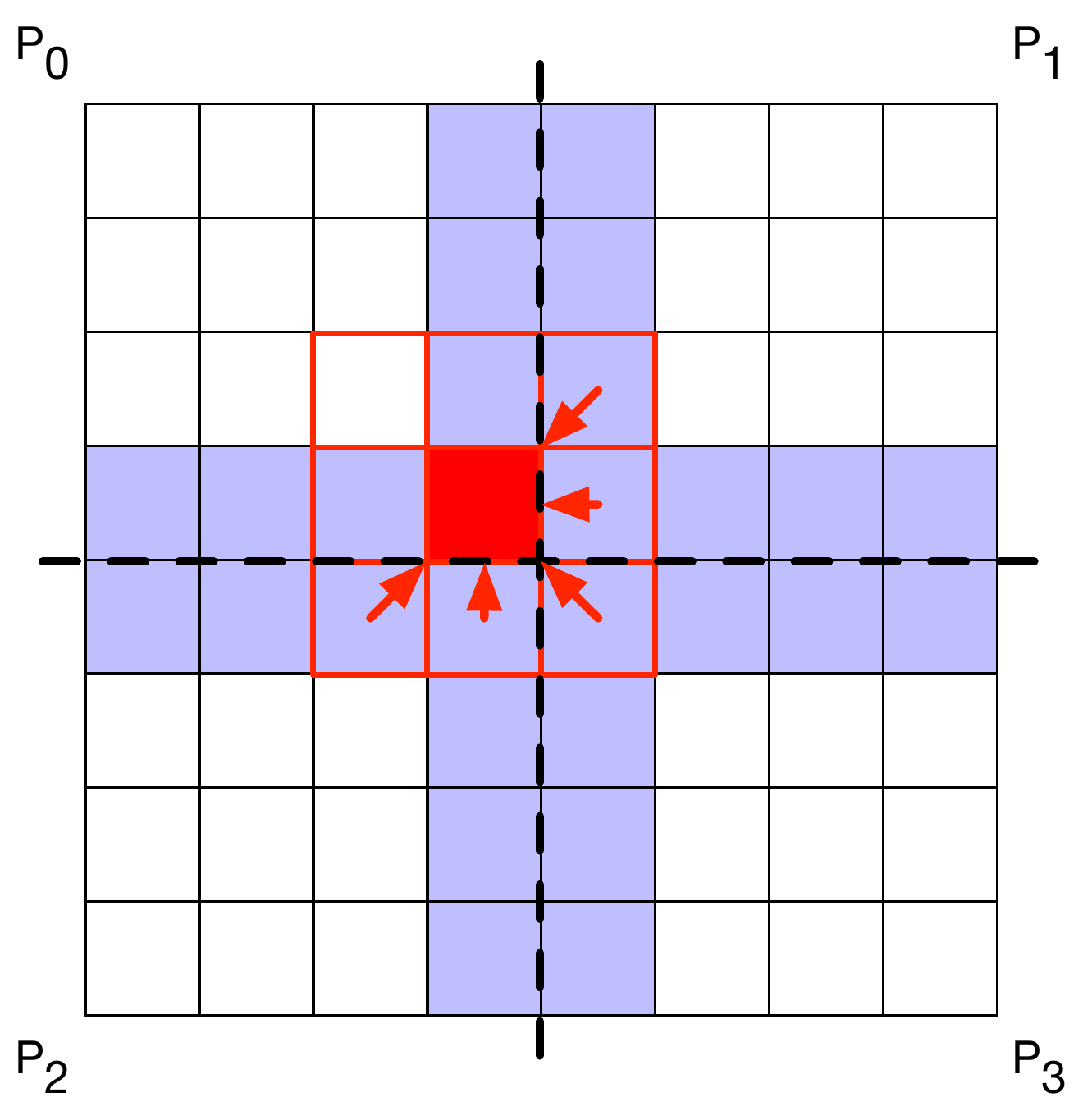}}
  \end{minipage}
  \caption{(a) Forward and backpropagation phases (red and blue arrows) for sample parallelism with two processors and a global mini-batch of size 4. (b) Example halo exchange for spatial parallelism on four processors. The solid red box is where the $3 \times 3$ convolution filter is centered and red arrows indicate data movement. The shaded region is the entire halo region.}
  \label{fig:algos}
\end{figure*}

At a high level, sample parallelism partitions $x$, $y$, $dL/dy$, and $dL/dx$ along the $N$ dimension, assigning complete samples to processors. The weights $w$ are stored redundantly on every processor. Computation of $y$ (forward propagation), $dL/dx$, and the local contributions of $dL/dw$ (backpropagation) can then be computed independently. An allreduce is required to complete the sum to form the final $dL/dw$ (the sum on $N$ in Eq~\ref{eq:bp-weights}), after which SGD can proceed independently on each processor.

Spatial parallelism is more complicated. The spatial dimensions $H$ and $W$ of $x$, $y$, $dL/dy$, and $dL/dx$ are split among processors. Most of forward propagation can be performed locally, but when a filter of size greater than $1 \times 1$ is placed near the border of a partition, remote data will be needed to compute the convolution. Thus a small number of rows and/or columns will need to be transferred from the remote processors in a halo exchange (as in a stencil computation). Backpropagation is similar, requiring a halo exchange on $dL/dy$ to compute $dL/dx$, and using the data from forward propagation to compute the local contributions of $dL/dw$. Finally, an allreduce completes the sum in $dL/dw$, like in sample parallelism. It should be observed that sample and spatial parallelism are orthogonal and can be used simultaneously. We refer to this as hybrid sample/spatial parallelism.

Figure~\ref{fig:algos} graphically illustrates sample parallelism and the halo exchange in spatial parallelism.

We now present our algorithm more formally, combining both sample and spatial parallelism. We keep the same simplifying assumptions as in Section~\ref{subsec:background-cnns}. Let $\distm{C}$ and $\distm{F}$ assign all indices to all processors (i.e. $w$ and $dL/dw$ are replicated on every processor). Let $\distm{N}$, $\distm{H}$, and $\distm{W}$ be distributions of the $N$, $H$, and $W$ dimensions. The size of the halo region will be $O$ rows or columns of length $\numelemd{W}$ or $\numelemd{H}$, respectively (we ignore padding issues here, but they are easy to handle). Let $q_H^{(p)} = \min \eleprocmd{H}$ be the lowest index in dimension $H$ assigned to processor $p$ and $r_H^{(p)} = \max \eleprocmd{H}$ the highest; define $q$ and $r$ similarly for $W$.

The non-local data in $x$ that processor $p$ requires (its halo) consists of the indices $( \eleprocmd{N}, \eleprocmd{C}, \{q_H^{(p)} - O, \ldots, q_H^{(p)} - 1, r_H^{(p)} + 1, \ldots, r_H^{(p)} + O\} \cup \eleprocmd{H}, \{q_W^{(p)} - O, \ldots, q_W^{(p)} - 1, r_W^{(p)} + 1, \ldots, r_W^{(p)} + O\} \cup \eleprocmd{W} ) \setminus \eleprocd$. Note that when $K=1$, $O=0$ and no halo is needed. Forward propagation on $p$ computes $y$ similarly to Eq~\ref{eq:fp}, except that the bounds on the indices for $y$ change: $k \in \eleprocmd{N}$, $f$ runs over all filters, $i \in \eleprocmd{H}$, and $j \in \eleprocmd{W}$. Notice that values of $x$ in the halo region will be required.

Backpropagation is adapted similarly. The halo region for $dL/dy$ is defined as for $x$, and the error signal $dL/dx$ is computed as in Eq~\ref{eq:bp-data}, where $k \in \eleprocmd{N}$, $c$ runs over all channels, $i \in \eleprocmd{H}$, and $j \in \eleprocmd{W}$. Computing the local gradients $dL/dw$ requires the halo region for $x$ and follows Eq~\ref{eq:bp-weights}:
\[ \frac{dL}{dw_{f,c,a,b}} = \sum_{k \in \eleprocmd{N}} \sum_{i \in \eleprocmd{H}} \sum_{j \in \eleprocmd{W}} \frac{dL}{dy_{k,f,i,j}} x_{k,c,i+a-O,j+b-O} \]
where $f$ and $c$ run over all filters and channels and $a$ and $b$ over $0, \ldots, K-1$. To compute the final value of $dL/dw$, an allreduce over all processors is performed. Note that $dL/dw$ does not require the halo region for $dL/dy$, and therefore the entire computation for $dL/dx$, including the halo exchange, can be performed concurrently with $dL/dw$.

In the case that the stride is greater than $1$, the process is similar, requiring that the halo region be adjusted to account for it. This may result in some processors not requiring a halo region. One important edge case to note is that spatial partitioning is complicated when a spatial dimension is the same size as the filter kernel. However, as spatial partitioning is most efficient when the kernel size (hence, halo area) is small compared to the partition size, as is typical, alternative parallelization approaches are preferred in this case.

\subsection{Entire CNNs}
\label{subsec:algo-entire}
The extension to an entire CNN is relatively straightforward. Input should be provided in the appropriate distribution for the first layer. Each convolutional layer can be parallelized as above. Pooling layers are parallelized similarly. Element-wise operations such as ReLUs parallelize trivially regardless of distribution. Batch normalization is typically computed locally on each processor; however, to our knowledge, performing batch normalization on subsets of the spatial dimensions has not been explored. Both purely local batch normalization and a variant that aggregates over the spatial distribution of a sample are easy to implement. FC layers use the model-parallel implementation of LBANN~\cite{van2015lbann}, although this can require a data redistribution, which we discuss next.

\subsection{Data redistribution}
\label{subsec:algo-redist}
It may occur that two layers $\ell_i$ and $\ell_j$ are adjacent, but use different data distributions $\dist_i$ and $\dist_j$ ($\eleproc{\dist_i} \neq \eleproc{\dist_j}$). This may be because it is more profitable to parallelize the layers using different approaches, or because one is a FC layer, which in our implementation uses an elemental distribution~\cite{poulson2013elemental}. When this occurs, data must be shuffled between the two distributions on both forward and backpropagation. This shuffle can be implemented via an all-to-all collective, where a processor sends indices it no longer owns ($\eleproc{\dist_i} \setminus \eleproc{\dist_j}$), and receives its new indices ($\eleproc{\dist_j} \setminus \eleproc{\dist_i}$).

\subsection{Channel and filter parallelism}
\label{subsec:algo-chanfilt}
We now briefly outline some approaches to channel and filter parallelism, although we leave implementation to future work. Channel and filter parallelism are closely related and data distributions must be chosen carefully; note, for example, if the input $x$ to a layer is partitioned on its $C$ dimension, the output $y$ is partitioned on its $F$ dimension. The choice of distributions significantly influences the communication involved, both within and between layers.

The computation of $y$ in forward propagation is relatively simple, except that the summation over channels ($c \in \eleprocmd{C}$) may involve a global reduce-scatter. Computation of $dL/dy$ is similar, with the summation over filters ($f \in \eleprocmd{F}$) possibly involving a global reduce-scatter. The gradients $dL/dw$ can be complicated to compute, as the update for a filter depends on both $x$ and $dL/dy$, which may require data to be gathered. A variety of data distributions are possible depending on the communication overheads and the amount of data replication that can be supported; we plan to explore these in future work.

\section{Implementation}
\label{sec:implementation}
In order to evaluate the proposed distributed convolution algorithms, we implement them by extending the LBANN toolkit for deep neural networks~\cite{van2015lbann}, which provides an underlying substrate for MPI-based parallel training with GPU acceleration. While LBANN can efficiently parallelize training using highly-optimized libraries such as cuDNN, NCCL, and Aluminum~\cite{Dryden2018}, it is limited to sample parallelism for convolution.

To extend LBANN for fine-grained distributed convolutions, we first develop a small C++ library for distributed tensor data structures that provides high-level abstractions for common tensor primitives used in CNN training. The design of the tensor library is strongly influenced by our prior experience in designing a high-level framework for stencil computations~\cite{Maruyama2011}.
It presents a partitioned global view of multidimensional tensors decomposed over distributed CPU and GPU memories.
For supporting convolutions, a halo exchange among adjacent distributed sub-tensors is implemented as part of the library API. It uses Aluminum for inter-node communication, and CUDA inter-process communication on-node. Most of these operations do not require the programmer to explicitly manage the data distribution. The library automates the underlying bookkeeping of distributed data structures as much as possible so long as this does not negatively impact performance. Fortunately, similar to regular stencil computations in scientific simulations, CNN computations tend to be rather regular, and thus realizing automation without performance penalties is possible (e.g. \cite{Maruyama2011} for stencils).

We implement a basic set of layers used in typical CNNs, including convolutions, pooling, batch normalization, and ReLU, on top of the tensor library. In this work, we focus on cluster systems with NVIDIA GPUs as the main computing platform for training and use NVIDIA's cuDNN library for optimized compute kernels. However, as cuDNN is not aware of the distribution of tensors, the library performs halo exchanges before convolutions and pooling.
We replace LBANN's tensor representation with ours with necessary data shuffling so that the overall training pipeline can be used as is.

\subsection{Optimization for strong scaling}

While fine-grained parallelization allows us to use a larger set of parallel resources, careful optimization of data movement becomes more important, especially for strong scaling. Unlike sample partitioning, halo exchanges are required in spatial partitioning and can be prohibitively expensive when the spatial domain is not large enough.

One of the well-known techniques for halo exchanges is overlapping with independent computations. Our implementation automatically decomposes an input tensor into its interior domain and boundary domains and calls cuDNN convolution kernels for each region separately so that halo exchanges can be run concurrently with the convolution of the interior domain. For backward convolutions, we exploit the task-level parallelism of backward data and filter convolutions to hide the halo exchange for the data convolution within the filter convolution. Note that the filter convolution does not require halo exchanges. We also minimize the latencies of halo exchanges by using asynchronous low-latency communication mechanisms such as GPUDirect RDMA if available.

\section{Performance model}
\label{sec:perfmodel}
We now provide a simple performance model for sample and spatial parallelism. Then we describe our approach for estimating good parallel execution strategies. All of this can be easily extended to handle channel and filter parallelism.

\subsection{Convolutional layer runtime}
\label{subsec:perf-conv}
We use empirical estimates for convolution, as cuDNN may select among many algorithms, and it is difficult to model GPU performance. These come from a simple benchmark that times the appropriate cuDNN function; we perform several warmup runs, then take the average of ten runs. This is combined with an analytic communication model (see Section~\ref{subsec:background-perfmodel}). Using this approach requires access to the target system in order to gather data, but does not require large-scale runs. As communication is often a chief bottleneck for training CNNs, an analytic model additionally allows flexibility to consider hypothetical communication optimizations.

We let $C(n, c, h, w, f)$ be the runtime of forward propagation (Eq.~\ref{eq:fp}) for $f$ filters on $n$ $h \times w$ samples with $c$ channels, with kernel size, stride, and padding omitted for convenience. This is used to estimate the \emph{local} runtime of convolution. Similarly, let $Cw(n, c, h, w, f)$ and $Cx(n, c, h, w, f)$ be the local runtimes for backpropagation (Eqs. \ref{eq:bp-weights} and \ref{eq:bp-data} respectively). For communication, we will use $AR(p, n)$ to denote the time for an allreduce of $n$ words over $p$ processors, and $SR(n)$ for the time to send and receive $n$ words between two processors. We separate these because allreduces use different algorithms (e.g., ring or butterfly) for different $n$ and $p$, so its performance cannot be directly deduced from point-to-point performance.

With this, the time for forward convolution for a layer $\ell$ is
\begin{align*}
  &FP_\ell = C(\numelemd{N}, \numelemd{C}, \numelemd{H}, \numelemd{W}, \numelemd{F}) \\
  &+ 2SR(O \numelemd{N} \numelemd{C} \numelemd{H}) + 2SR(O \numelemd{N} \numelemd{C} \numelemd{W}) \\
  &+ 4SR(O^2 \numelemd{N} \numelemd{C})
\end{align*}
where the send/recvs are for the east/west, north/south, and corner halo regions, respectively. If $\numelemd{W} = W$, then the east/west and corner halo exchanges can be omitted; if $\numelemd{H} = H$, then the north/south and corner halo exchanges can be omitted. Additionally, if the implementation supports it, the halo exchanges can be overlapped with interior computation.

For backpropagation, the time to compute $dL/dx$ is
\begin{align*}
  &BPx_\ell = Cx(\numelemd{N}, \numelemd{C}, \numelemd{H}, \numelemd{W}, \numelemd{F}) \\
  &+ 2SR(O \numelemd{N} \numelemd{C} \numelemd{H}) + 2SR(O \numelemd{N} \numelemd{C} \numelemd{W}) \\
  &+ 4SR(O^2 \numelemd{N} \numelemd{C}).
\end{align*}
Note the similarity to forward propagation; the same discussion on halo exchanges and overlapping applies here. We will simply write $BPw_\ell = Cw(\numelemd{N}, \numelemd{C}, \numelemd{H}, \numelemd{W}, \numelemd{F})$ to be the time to compute the \emph{local} portion of $dL/dw$. Finally, the allreduce time to complete $dL/dw$ is $BPa_\ell = AR(|\procsp{\distm{C}, \distm{F}}|, \numelemd{F} \numelemd{C} K^2)$. These are written separately because there are several additional, commonly used opportunities for overlap. The halo exchanges for computing $dL/dx$ can additionally be overlapped with computing $dL/dw$; both local computations could be done concurrently; and the allreduce can be overlapped with computing $dL/dx$, as well as computations in other layers.

We will write $\Costl{\dist}{\ell_i} = FP_{\ell_i} + BPx_{\ell_i} + BPw_{\ell_i} + BPa_\ell$ to be the cost of a layer under distribution $\dist$, adjusting for overlap if necessary.

From these performance models, we can see that in terms of communication overheads, sample parallelism is the ``cheapest'' approach: it requires only the allreduce time in $BPa_\ell$. Spatial parallelism adds additional overhead from the halo exchanges (which can be overlapped).

\subsection{Runtime for a CNN}
\label{subsec:perf-cnn}
Extending this to an entire CNN is relatively straightforward. We need to address three things: layers besides convolution, overlapping between layers, and data redistributions. As most layers other than convolution and FC layers are computationally cheap, we treat them as free, for simplicity. FC layers use the standard performance models for distributed matrix multiplication~\cite{schatz2016parallel}. If a layer has learnable parameters (e.g. batch normalization), an allreduce is required, and modeled similarly to above (note that model-parallel FC layers do not need such an allreduce).

We estimate allreduce overlap between layers (e.g. $BPa$) by greedily overlapping as much computation as possible with an allreduce. Only one allreduce at a time is considered to run, which neglects the possibility of running multiple concurrently, but simplifies the model.

Data redistribtion may be required in forward and backpropagation when the distributions of adjacent layers differ (Section~\ref{subsec:algo-redist}). This uses an all-to-all to shuffle data; we use $\Shuffle{\dist_i}{\dist_j}$ for the cost of moving data from $\dist_i$ to $\dist_j$.

\subsection{Parallel execution strategies}
\label{subsec:perf-strats}
It may be advantageous to use different distributions of data for different layers in a CNN, in order to parallelize them differently. For example, spatial parallelism is unlikely to provide a significant benefit to a layer with a small spatial domain. This introduces a number of knobs that can be tuned to parallelize a network, and selecting the appropriate parallelization scheme for each layer is difficult. Further, the performance of convolution implementations can be hard to predict in advance. A \emph{parallel execution strategy} for a network is an assignment of distributions to each layer. We now sketch a simple optimization approach for selecting good parallel execution strategies using our performance model.

First we generate candidate distributions for each layer. For convolutional layers, we heuristically select distributions that are load balanced and prefer cheaper partitioning methods (i.e. sample over spatial parallelism) when possible. We assume FC layers are either entirely sample- or model-parallel and that other layers simply inherit the distribution of their parent.

Given the set of candidate distributions for each layer, we find the best parallel execution strategy for the CNN by reducing to the single-source shortest path problem. For simplicity, we first consider the case where the CNN has a ``line'' architecture with no branches. We construct a graph with a vertex for every candidate distribution for every layer, plus source and sink vertices. There is an edge from each candidate distribution $\dist_i$ for a layer $\ell_i$ to every candidate distribution $\dist_j$ of its child layer $\ell_j$, weighted with $\Costl{\dist_i}{\ell_i} + \Shuffle{\dist_i}{\dist_j}$. Finally, there is an edge from the source vertex to every candidate distribution of the first layer with weight 0, and from every candidate distribution of the last layer $\ell_i$ to the sink vertex weighted by $\Costl{\dist_i}{\ell_i}$.

A shortest path from the source to the sink of this graph gives a parallel execution strategy with the fastest end-to-end runtime for the CNN. Since this is a directed acyclic graph, this path can be found in linear time. While a large number of vertices and edges are generated, we have found this is not an issue in practice. If necessary, additional heuristics could be used to prune them.

For networks that have branches (e.g. ResNets), we cannot directly apply this approach, as some layers have multiple parents or children. Instead, we first find the longest path from the beginning to the end of the CNN, and apply the above to every layer in this path. The distributions for these layers are fixed, and we repeat with the next longest path that contains as few of the already-used layers as possible until every layer has a distribution. The idea behind this approach is that the longest path is the most computationally intensive, and so it should be optimized first, to guarantee maximum flexibility in distribution choice.

\section{Evaluation}
\label{sec:eval}
We now evaluate our algorithms via microbenchmarks and end-to-end training. We use Lassen, a CORAL-class supercomputer~\cite{lassen}, which consists of 650 nodes, each with two IBM POWER9 CPUs and four Nvidia V100 (Volta) GPUs with NVLINK2 and 16 GB of memory per GPU, interconnected via dual-rail InfiniBand EDR. Our implementation uses a recent development version of LBANN and Aluminum. We use GCC 7.3.1, Spectrum MPI 2019.01.30, CUDA 9.2.148, cuDNN 7.4.1, and NCCL 2.4.2.

We consider two networks: a fully-convolutional ResNet-50~\cite{he2016deep,long2015fully} for ImageNet-1K classification and a proof-of-concept model for a 2D mesh-tangling problem which we formulate as semantic segmentation. The data consists of images representing a hydrodynamics simulation state at a timestep, and the problem is to predict, for each pixel, whether the mesh cell at that location needs to be relaxed to prevent tangling. Mesh tangling occurs when cells overlap, which is non-physical and results in the simulation degenerating or failing entirely. It is hoped that CNNs, which can incorporate global information from the simulation state, will offer better heuristics for predicting tangling than existing methods. However, as interesting simulations are high-resolution, it has not been possible to train neural networks on the data due to memory constraints.

For these tests, the input data is either $1024 \times 1024$ ($\MeshI$) or $2048 \times 2048$ ($\MeshII$) pixel images, with 18 channels consisting of various state variables and mesh quality metrics from a hydrodynamics simulation. We use 10,000 samples of each size. Our CNN is a very simple fully-convolutional model adapted from VGGNet~\cite{simonyan2014very} for our input sizes and semantic segmentation. It consists of six blocks of either three ($\MeshI$) or five ($\MeshII$) convolution-batch normalization-ReLU operations, using $3 \times 3$ convolutional filters, and a final convolutional layer for prediction. Downsampling is performed via stride-2 convolution at the first convolutional filter of each block. The model for the $\MeshII$ mesh data is large enough, when including intermediate activations, to exceed GPU memory when training with even one sample. For performance benchmarks on this problem, we use synthetic data, as our goal is to focus on the performance of our algorithms and demonstrate that models of this scale can be quickly trained on HPC systems. We leave developing optimized models to future work, now that training is feasible.

\subsection{Layer benchmarks}
\label{subsec:eval-bm}

\begin{figure*}
  \centering
  \begin{minipage}{0.49\textwidth}\centering
    \subfloat
    {\includegraphics[scale=0.55]{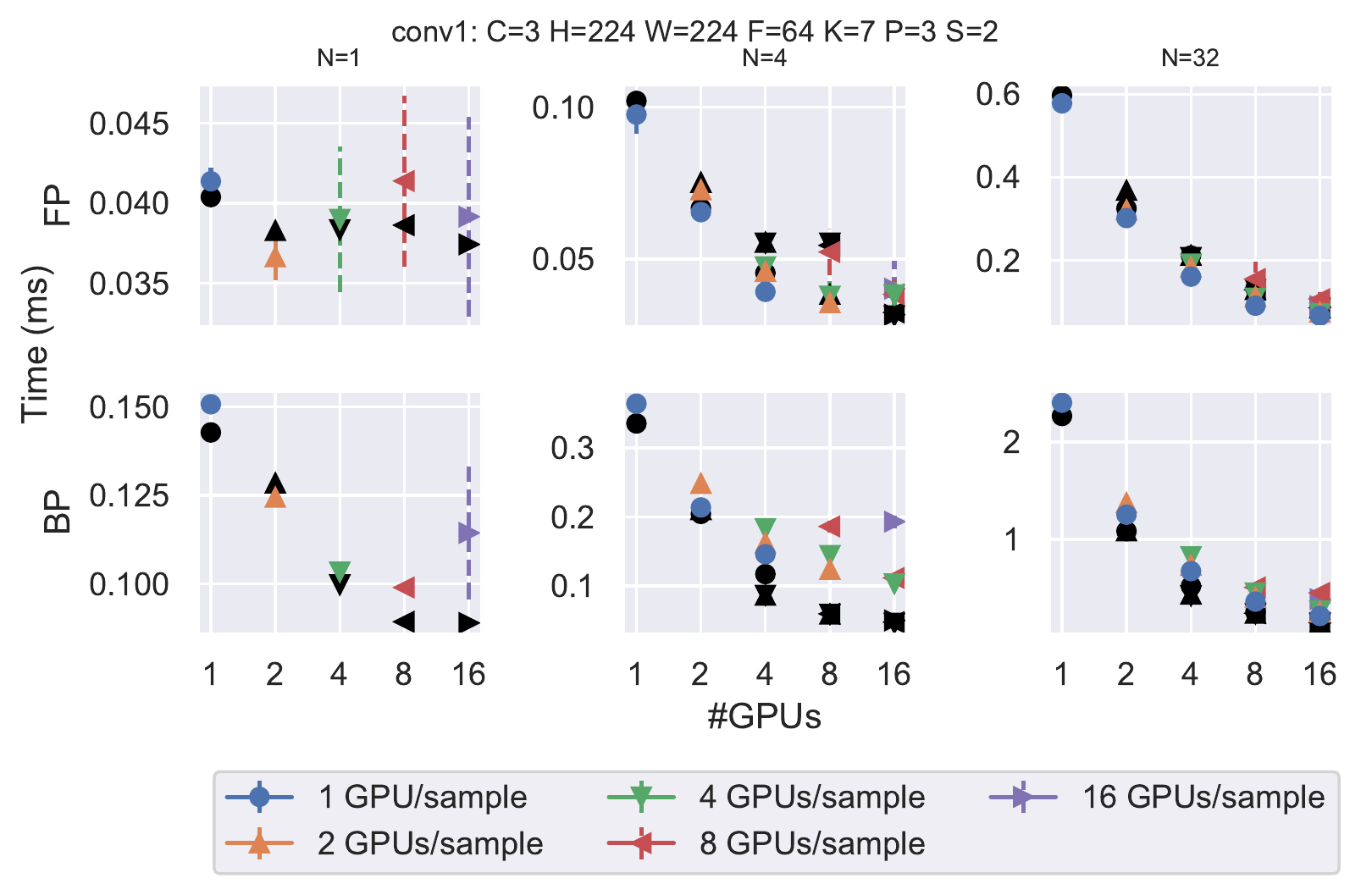}}
  \end{minipage}
  \begin{minipage}{0.49\textwidth}\centering
    \subfloat
    {\includegraphics[scale=0.55]{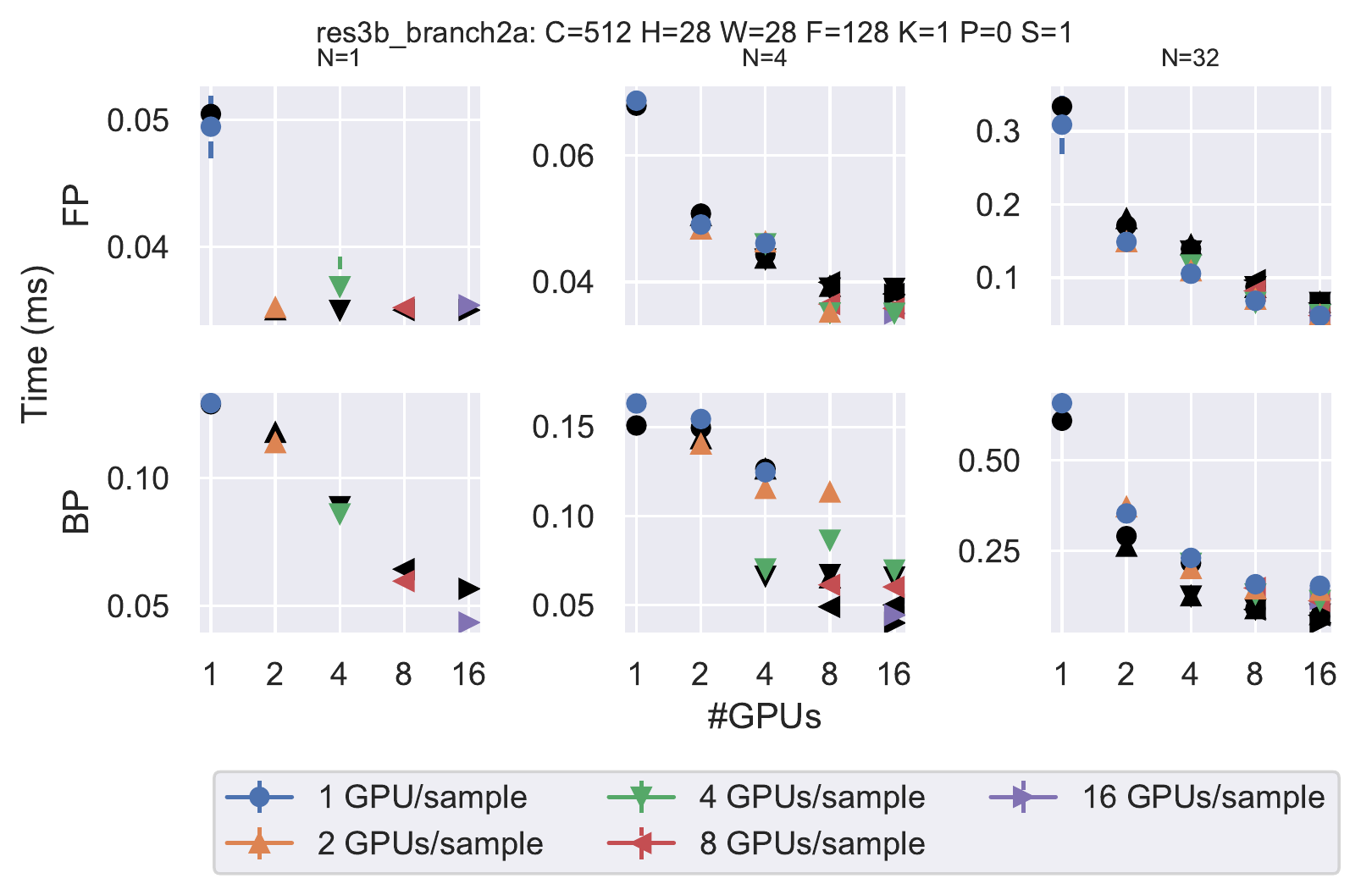}}
  \end{minipage}
  \caption{Microbenchmark results for layers \texttt{conv1} and \texttt{res3b\_branch2a} of ResNet-50 comparing parallelization schemes in forward (FP) and backpropagation (BP). Error bars are $\pm$ one standard deviation. Black shapes are performance model predictions. Specifications for each layer are above each figure.}
  \label{fig:microbm-resnet}
\end{figure*}

\begin{figure*}
  \centering
  \begin{minipage}{0.49\textwidth}\centering
    \subfloat
    {\includegraphics[scale=0.55]{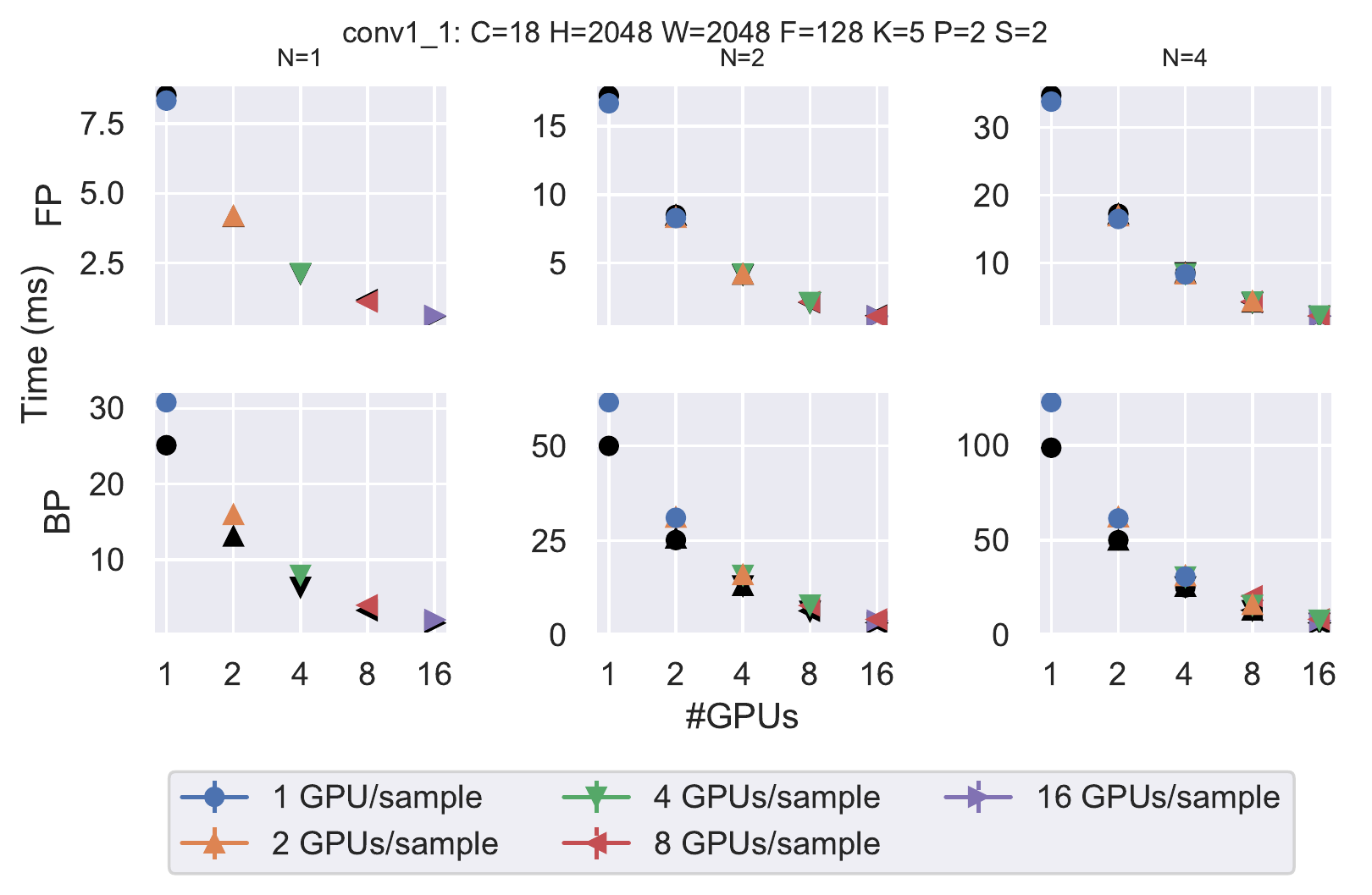}}
  \end{minipage}
  \begin{minipage}{0.49\textwidth}\centering
    \subfloat
    {\includegraphics[scale=0.55]{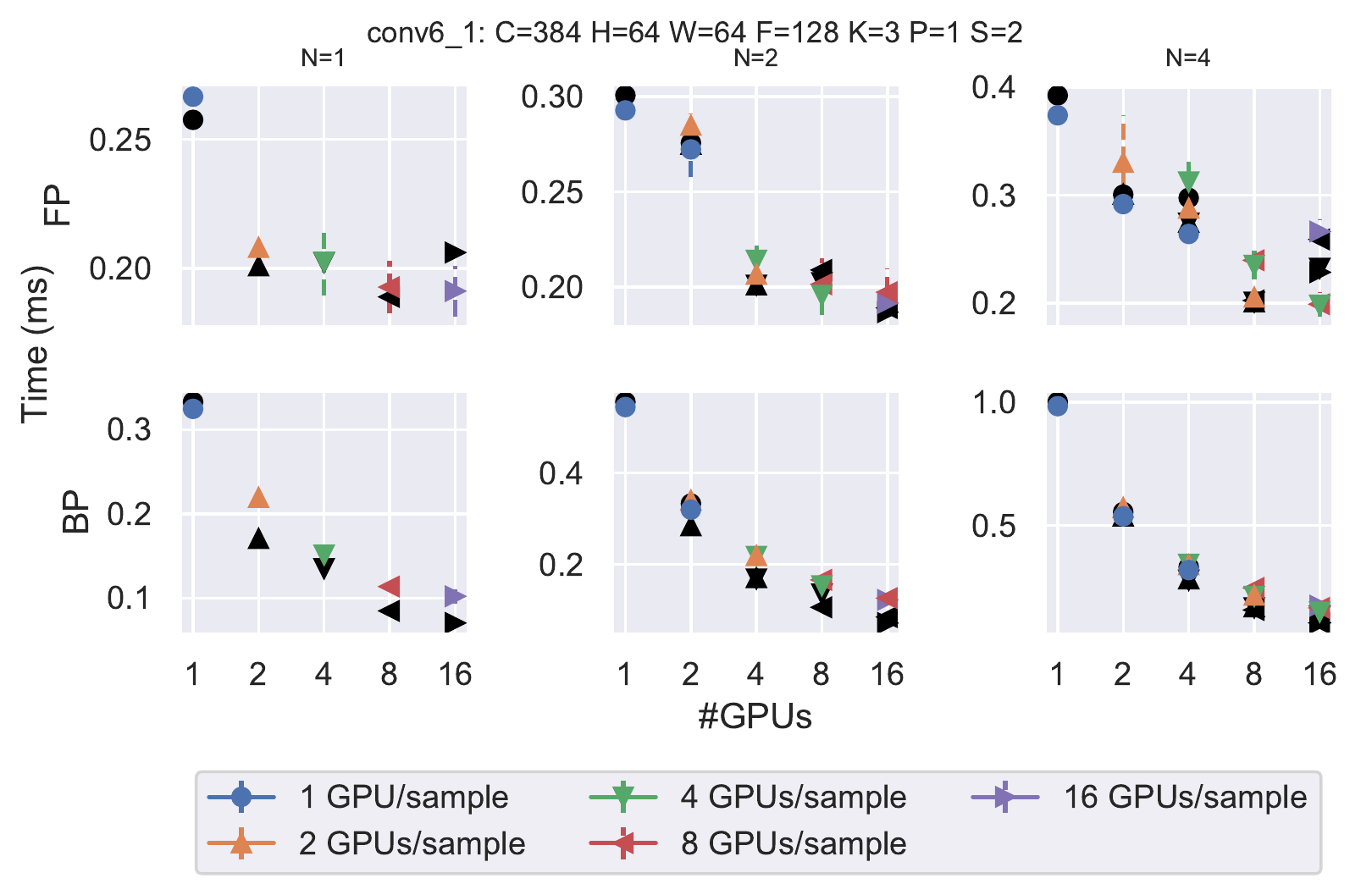}}
  \end{minipage}
  \caption{Microbenchmark results for layers \texttt{conv1\_1} and \texttt{conv6\_1} from the $\MeshII$ mesh model, comparing parallelization schemes in forward (FP) and backpropagation (BP). Error bars are $\pm$ one standard deviation. Black shapes are performance model predictions. Specifications for each layer are above each figure.}
  \label{fig:microbm-mesh}
\end{figure*}

We first present microbenchmark results for selected layers from ResNet-50 and the $\MeshII$ mesh model on up to four Lassen nodes. These enable the performance characteristics of spatial convolution to be seen at a fine-grained scale. We time forward and backpropagation of each layer, with halo exchanges being overlapped. We exclude the allreduce to accumulate gradients to focus on the performance of convolution. Also, it is typically overlapped by other computation. For each measurement, we first do warmup runs, then report the mean and standard deviation of ten runs.

Figure~\ref{fig:microbm-resnet} shows results for ResNet-50 layers \texttt{conv1} and \texttt{res3b\_branch2a} with $N=$ 1, 4, and 32 samples. The first two can occur when strong scaling sample-parallelism to few samples per GPU; the last is a typical target for efficient use of a GPU. \texttt{conv1} is the first layer of ResNet, with a relatively large ($224 \times 224$) input, but only three channels and 64 filters. The kernel is large, $K=7$, requiring large halo exchanges for spatial parallelism. For $N=1$, forward propagation does not scale well, due to limited computation to hide halo exchanges; backpropagation fares better, and results in net improvements to forward and backpropagation of ${\sim}1.35$x with 8 GPUs. Performance degrades somewhat with 16 GPUs, due to communication overheads. \texttt{res3b\_branch2a} is a $1 \times 1$ convolution from the middle of ResNet, with a fairly small spatial domain. The filter size means that no halo exchange is needed, avoiding communication overheads. Forward propagation does not show significant performance improvements beyond two GPUs, due to fixed kernel overheads. Backpropagation shows improvements up to 16 GPUs except that the 2 GPUs/sample case is significantly slower than 4 GPUs/sample at 4 GPUs due to the performance of the underlying cuDNN kernels. With larger numbers of samples, spatial decomposition remains competitive with pure sample parallelism, indicating halo exchanges are hidden.

Figure~\ref{fig:microbm-mesh} presents results for two layers of our $2048 \times 2048$ mesh model. Here spatial domains are much larger, and we expect spatial parallelism to perform better. Results are for $N=$ 1, 2, and 4 samples, since due to the size of the data, only one or two samples can be trained per node. (Since we benchmark only individual layers here, memory pressure is not as significant as with an entire network.) \texttt{conv1\_1} is the first layer, and has extremely large spatial input. The $N=1$ case has very good scaling on both forward and backpropagation, achieving ${\sim}14.8$x speedup on 16 GPUs, indicating inter-node halo exchange overheads are well-hidden. The two sample case is similar. With four samples, the overhead of the halo exchange is very minor, enabling competitive scaling to sample parallelism. \texttt{conv6\_1} is from much deeper within the network, and has a smaller input spatial domain. Nonetheless we see continued benefit in the $N=1$ case (${\sim}1.4$x).

These results illustrate several important things. First, that the empirical performance of convolution can be complicated to predict. Second, that the intuition from our performance model is broadly correct: other things equal, sample parallelism typically has the least overhead. Finally, the small $N$ case, observed when strong-scaling sample-parallelism, can benefit significantly from spatial parallelism.

\subsection{Training}
\label{subsec:eval-training}

\begin{figure*}
  \centering
  \begin{minipage}{0.49\textwidth}\centering
    \subfloat
    {\includegraphics[scale=0.56]{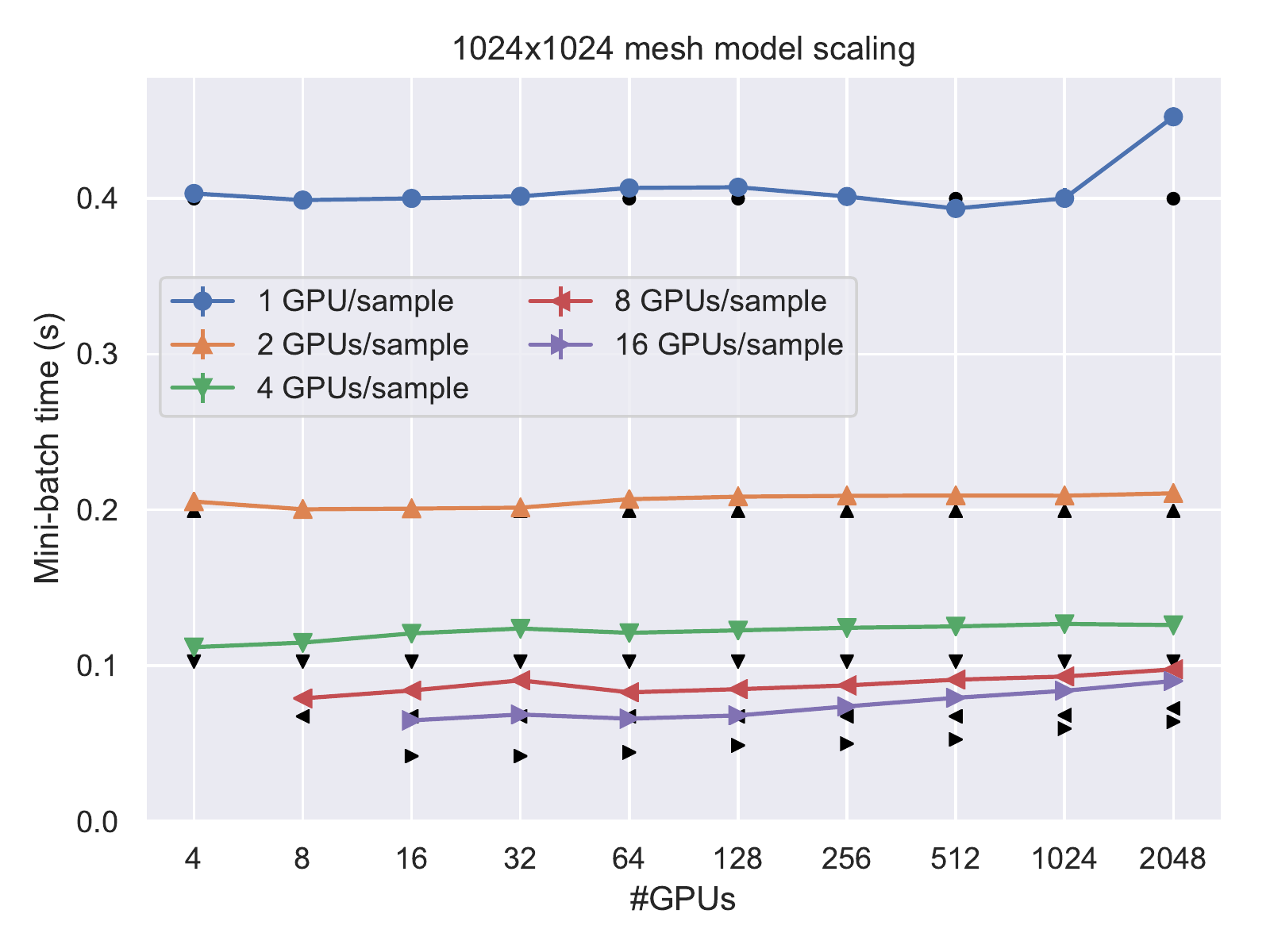}}
  \end{minipage}
  \begin{minipage}{0.49\textwidth}\centering
    \subfloat
    {\includegraphics[scale=0.56]{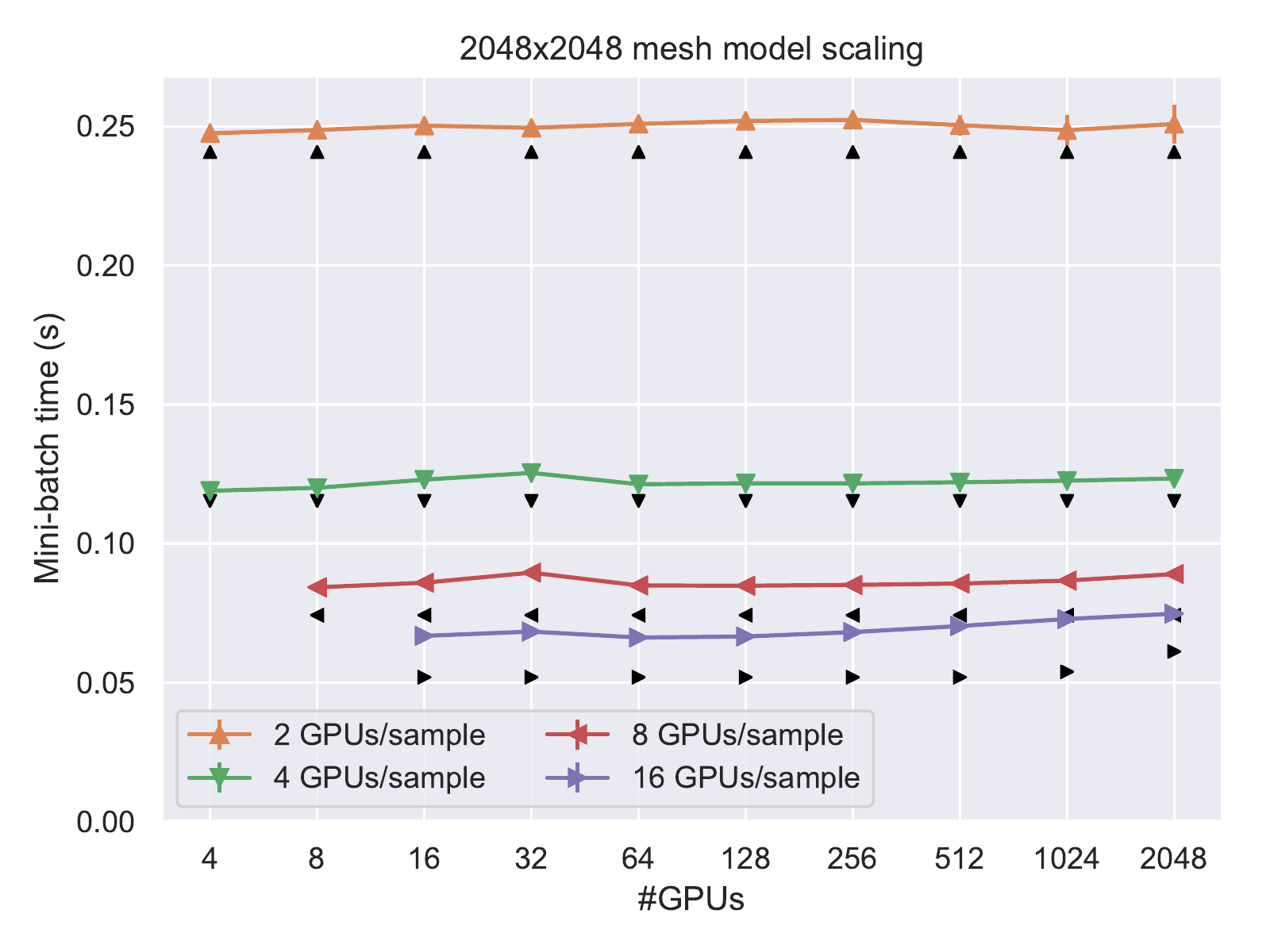}}
  \end{minipage}
  \caption{Weak scaling for the $\MeshI$ and $\MeshII$ mesh models. Error bars are $\pm$ one standard deviation. Black markers are estimates from our performance model. 1 sample/GPU is is pure sample parallelism; other cases are hybrid sample/spatial parallelism. Note that there are 4 GPUs/node. Spatial parallelism is required for the $\MeshII$ model due to memory requirements. Strong scaling can be seen between parallelization schemes (see Tables~\ref{tab:mesh1k-speedup} and \ref{tab:mesh2k-speedup}).}
  \label{fig:mesh-scaling}
\end{figure*}

We now present scaling results for end-to-end training of our models. We consider both strong and weak scaling, although our focus is strong scaling. We define strong scaling as parallelizing a CNN over more GPUs, while keeping the mini-batch size and all other aspects fixed. Weak scaling parallelizes a CNN by keeping the mini-batch size per GPU fixed, and growing the global mini-batch size as more GPUs are added, working on the same problem but with larger batch sizes. Strong scaling has the advantage that one need not address issues of learning and generalization with large mini-batches, as the learning process does not change: once a good mini-batch size is selected, it can be strong-scaled to make training faster. Our results are primarily hybrid sample-spatial parallelism, where samples are first partitioned onto groups of GPUs, and then spatially parallelized within that group. We use the same data decomposition for every layer in a given configuration, although this is not necessarily optimal; we leave exploring more varied decompositions to future work.

\subsubsection{Mesh model}

\begin{table*}[h]
\caption{$\MeshI$ mesh strong scaling at mini-batch size $N$, mini-batch time and speedup over 1 GPU/sample (sample parallelism).}
\centering
\renewcommand{\arraystretch}{1.2}
\begin{tabular}{llllll}
\toprule
$N$ & 1 GPU/sample & 2 GPUs/sample & 4 GPUs/sample & 8 GPUs/sample & 16 GPUs/sample \\
\midrule
4 & 0.403s & 0.2s (2.0x) & 0.121s (3.3x) & 0.0906s (4.4x) & 0.066s (6.1x) \\
8 & 0.399s & 0.201s (2.0x) & 0.124s (3.2x) & 0.0829s (4.8x) & 0.0681s (5.9x) \\
16 & 0.4s & 0.201s (2.0x) & 0.121s (3.3x) & 0.085s (4.7x) & 0.0739s (5.4x) \\
32 & 0.401s & 0.207s (1.9x) & 0.123s (3.3x) & 0.0874s (4.6x) & 0.0794s (5.1x) \\
64 & 0.407s & 0.208s (2.0x) & 0.124s (3.3x) & 0.0911s (4.5x) & 0.0839s (4.8x) \\
128 & 0.407s & 0.209s (1.9x) & 0.125s (3.3x) & 0.0931s (4.4x) & 0.0902s (4.5x) \\
256 & 0.401s & 0.209s (1.9x) & 0.127s (3.2x) & 0.0977s (4.1x) & n/a \\
512 & 0.393s & 0.209s (1.9x) & 0.126s (3.1x) & n/a & n/a \\
1024 & 0.4s & 0.211s (1.9x) & n/a & n/a & n/a \\
\bottomrule
\end{tabular}
\label{tab:mesh1k-speedup}
\end{table*}

\begin{table*}[h]
\caption{$\MeshII$ mesh strong scaling at mini-batch size $N$, mini-batch time and speedup over 2 GPUs/sample.}
\centering
\renewcommand{\arraystretch}{1.2}
\begin{tabular}{llllll}
\toprule
$N$ & 2 GPUs/sample & 4 GPUs/sample & 8 GPUs/sample & 16 GPUs/sample \\
\midrule
2 & 0.247s & 0.12s (2.1x) & 0.0859s (2.9x) & 0.0683s (3.6x) \\
4 & 0.249s & 0.123s (2.0x) & 0.0895s (2.8x) & 0.0662s (3.8x) \\
8 & 0.25s & 0.125s (2.0x) & 0.0849s (2.9x) & 0.0665s (3.8x) \\
16 & 0.249s & 0.121s (2.1x) & 0.0848s (2.9x) & 0.0681s (3.7x) \\
32 & 0.251s & 0.122s (2.1x) & 0.0851s (2.9x) & 0.0703s (3.6x) \\
64 & 0.252s & 0.122s (2.1x) & 0.0856s (2.9x) & 0.0729s (3.5x) \\
128 & 0.252s & 0.122s (2.1x) & 0.0867s (2.9x) & 0.0748s (3.4x) \\
256 & 0.25s & 0.123s (2.0x) & 0.089s (2.8x) & n/a \\
512 & 0.249s & 0.123s (2.0x) & n/a & n/a \\
\bottomrule
\end{tabular}
\label{tab:mesh2k-speedup}
\end{table*}

Figure~\ref{fig:mesh-scaling} shows scaling results for the $\MeshI$ and $\MeshII$ mesh models, up to 2048 GPUs (512 nodes). For the former, we present five cases: sample parallelism, and 2-, 4-, 8-, and 16-way hybrid sample/spatial parallelism with mini-batch sizes $N = 4$ to $2048$ (omitting cases when they require too many GPUs). The model can fit only one sample per GPU, so we do not explore additional sample parallelism. We run the same configurations for the $\MeshII$ mesh model, except pure sample parallelism is not possible due to memory constraints and our maximum mini-batch size is $1024$. We compare strong-scaling at a fixed mini-batch size across parallelism cases, and weak scaling as the mini-batch size grows. Note that when using 8- or 16-way spatial parallelism, a sample is being partitioned across two or four nodes, requiring both intra- and inter-node communication for halo exchanges.

For strong scaling, spatial parallelism allows us to use additional GPUs for the same mini-batch size. Table~\ref{tab:mesh1k-speedup} gives mini-batch times and speedups for the $\MeshI$ mesh model. We see near-linear speedup for 2 GPUs/sample over sample parallelism (2x is ideal), and significant further improvements with 4 GPUs/sample. Improvements continue with 8 and 16 GPUs/sample, although they are not as dramatic due to the increased overheads of halo communication and local convolution kernels not scaling linearly. Weak scaling results can be seen in Figure~\ref{fig:mesh-scaling} (left), where the flat mini-batch time for increasing numbers of GPUs (hence, increasing mini-batch size) shows near-perfect weak scaling. This implies that our spatial partitioning is not impacting the typical sample-parallel weak scaling trends. Weak scaling for 8 and 16 GPUs/sample does show a slight trend of increasing mini-batch time at large scale; due to the extensive data decomposition, each GPU has significantly less work, and our implementation cannot fully overlap global allreduces with backpropagation computation in these cases.

The performance degradation for sample parallelism at 2048 GPUs is due to memory pressure requiring a smaller workspace for cuDNN, impacting local convolution algorithm selection. The increased memory pressure is due to communication-related data structures taking increased GPU memory, and could be mitigated with future optimizations.

Results for the $\MeshII$ mesh model are shown in Table~\ref{tab:mesh2k-speedup} (strong scaling) and Figure~\ref{fig:mesh-scaling} (right, weak scaling). The strong scaling trend is similar to that for the $\MeshI$ model, with comparable speedups from each algorithm. Indeed, when increasing from 4 GPUs/sample to 8 GPUs/sample, the models observe roughly a $1.3x$ and $1.4x$ improvement in performance (respectively). Neither observe linear increases due to the high overhead of fine-grained inter-node halo communication, but our overlapping enables further improvement nonetheless. The $\MeshII$ model achieves slightly more speedup in this case, as there is more work to overlap communication with. The weak scaling trend is also comparable to that for the $\MeshI$ model, although we only observe weak scaling performance degrading with 16 GPUs/sample for the $\MeshII$ model.

That we can achieve both good strong and weak scaling for the mesh-tangling problem is important for being able to rapidly train and explore new models: good strong scaling means we can accelerate training of models without changing learning dynamics, and good weak scaling additionally helps with tuning mini-batch sizes.

\subsubsection{ResNet-50}

\begin{table*}[h]
\caption{Strong scaling ResNet-50, mini-batch time and speedup over sample parallelism.}
\centering
\renewcommand{\arraystretch}{1.2}
\begin{tabular}{llll}
\toprule
$N$ & Sample (32 samples/GPU) & Hybrid (32 samples/2 GPUs) & Hybrid (32 samples/4 GPUs) \\
\midrule
128 & 0.106s & 0.0734s (1.4x) & 0.0593s (1.8x) \\
256 & 0.106s & 0.0732s (1.4x) & 0.0671s (1.6x) \\
512 & 0.105s & 0.0776s (1.4x) & 0.0617s (1.7x) \\
1024 & 0.105s & 0.0747s (1.4x) & 0.0672s (1.6x) \\
2048 & 0.108s & 0.0733s (1.5x) & 0.0651s (1.7x) \\
4096 & 0.0984s & 0.078s (1.3x) & 0.066s (1.5x) \\
8192 & 0.109s & 0.0785s (1.4x) & 0.0725s (1.5x) \\
16384 & 0.108s & 0.0844s (1.3x) & 0.0792s (1.4x) \\
32768 & 0.109s & 0.0869s (1.3x) & n/a \\
\bottomrule
\end{tabular}
\label{tab:resnet-speedup}
\end{table*}

We present strong scaling results for ResNet-50, comparing pure sample parallelism to hybrid sample 2-way and 4-way parallelism, in Table~\ref{tab:resnet-speedup}. We use 32 samples per GPU as our baseline, as this is a typical choice to saturate GPUs. Using spatial parallelism we achieve 1.4x speedups with 2x as many GPUs, and up to 1.8x with 4x as many GPUs. Prior work has shown that strong scaling this problem size via sample-parallelism past $\sim$8-16 nodes rapidly results in communication overhead making it unprofitable~\cite{Dryden2018}.

Thus, weak scaling is typically preferred, ensuring sufficient local work to hide communication costs. However, one cannot weak scale indefinitely, even with large mini-batch techniques, due to generalization issues. Goyal et al.~\cite{goyal2017accurate} report ImageNet validation error begins to increase with mini-batch sizes above 8k when using ResNet-50. For different problems, datasets, or network architectures, the max mini-batch size will be different~\cite{shallue2018measuring}. To continue to accelerate training beyond this, strong scaling must still be employed. Table~\ref{tab:resnet-speedup} shows we get continued improvement from spatial parallelism with larger mini-batch sizes. Speedups decrease slightly at larger scale for more extensive decomposition, due to the implementation being unable to fully overlap the cost of allreduces.

Achieving near-linear speedup for ResNet is unlikely, as most layers have small spatial domains. This agrees with our microbenchmarks. (Channel/filter parallelism may be more promising, as many layers have many filters.) Despite this, we are still able to accelerate many problem sizes of interest.

\subsubsection{Performance model}
Our figures in this section have also included the corresponding predictions from our performance model. We can see that its predictions are quite accurate, and even when there are deviations, it still has the correct trend and ranking of algorithms. Much of the inaccuracy is due to lower-order computations that are not accounted for but matter with more extensive decompositions (e.g. 16 GPUs/sample). Network noise and other similar factors are also not accounted for. Nevertheless, this validates the accuracy of the performance model, and we can be confident that generating parallel execution strategies with it will be effective.

\section{Related Work}
\label{sec:related}
There are many techniques for parallelizing CNNs at different scales. Ben-Nun and Hoefler~\cite{ben2018demystifying} provide a comprehensive overview of approaches. Recent work has leveraged HPC resources for training CNNs on large simulation data (e.g. \cite{kurth2017deep,kurth2018exascale,mathuriya2018cosmoflow}). These works have primarily focused on scaling sample-parallel training via optimizing communication and I/O, large mini-batches, etc. They do not address the issues of very large samples. These optimizations are orthogonal to our algorithms, and they can be used together.

\textbf{Scaling convolution.} AlexNet~\cite{krizhevsky2012imagenet} used an early form of model parallelism, where convolutional filters and fully-connected layers were partitioned between two GPUs to avoid memory limits. In this implementation, grouped convolutions were used for certain layers to reduce inter-GPU communication costs, instead of directly replicating the result of regular convolution. Coates et al~\cite{coates2013deep} used a distributed tensor representation to spatially partition locally-connected layers; we use similar techniques for convolutional layers. Gholami et al.~\cite{gholami2017integrated} also propose general decompositions of convolutional layers, but only provide simulated results. Some deep learning frameworks, such as DistBelief~\cite{dean2012large} and Project Adam~\cite{chilimbi2014project} support partitioning the filters of convolutional layers onto workers and aggregating their results via parameter server. Oyama et al.~\cite{oyama2018accelerating} provide a framework to split mini-batches into smaller batches to use faster convolution algorithms.

Demmel and Dinh~\cite{demmel2018communication} give lower bounds on communication for convolutional layers and sequential algorithms that achieve them. However, their analysis is limited to forward propagation of a single layer, and does not consider optimizing training time for an entire CNN.

Large mini-batch training has been developed via linear scaling of learning rates~\cite{goyal2017accurate} or layer-wise adaptive learning rates~\cite{you2018imagenet}. These approaches have been applied primarily to image classification, and can require extensive hyperparameter tuning. These are complementary to our work: when viable, large mini-batches can enable large-scale sample-parallel training, but we provide additional options for parallelization.

\textbf{Parallel execution strategies.} Yan et al.~\cite{yan2015performance} develop a performance model for training DNNs with Adam~\cite{chilimbi2014project} and an optimizer for finding good configurations. Their system does not include spatial partitioning, and their performance was evaluated on a CPU system, limiting its applicability to modern training regimes. Mirhoseini et al.~\cite{mirhoseini2017device} use reinforcement learning for TensorFlow graph placement on a small heterogeneous system, but do not consider partitioning convolutions beyond the sample dimension.

\textbf{Memory pressure.} Several approaches to alleviating memory pressure on GPUs have been used. If at least one sample can fit in GPU memory, an out-of-core ``micro-batching'' approach, where mini-batches are split into micro-batches and updates accumulated, can be used, but this can increase training time~\cite{ito2017ooc_cudnn}. Other approaches utilize recomputation to avoid keeping intermediate values~\cite{chen2016training}. ``Prefetching'' and dynamic memory movement approaches, either through CUDA unified memory or in training frameworks~\cite{rhu2016vdnn,meng2017training,wang2018superneurons}, can be used, but this can require prefetch hints and result in additional CPU/GPU memory transfers, reducing the bandwidth available for communication.

\section{Conclusions}
\label{sec:conclusions}
We have demonstrated additional approaches for extracting parallelism from convolutional layers for training CNNs, enabling improved strong scaling. Exploiting parallelism within the spatial domain allows scaling to continue beyond the mini-batch size. This is necessary to train models on datasets with very large samples, which would otherwise be infeasible due to memory constraints, and enables further acceleration of training once the limits of sample-parallelism have been reached. As scientific domains and HPC simulations take greater advantage of deep learning, handling massive samples will only become more important. In particular, as 3D data becomes more widespread, spatial parallelism, which can be easily extended to 3D, becomes critical, and more advantageous, due to the more favorable surface-to-volume ratio.

We have also presented a performance model that reasonably approximates observed performance and provides intuition about different approaches to parallelizing convolution. Leveraging this to select good parallel execution strategies for CNNs will only become more important as they grow more complex and additional parallelism is exploited.

There remains significant additional sources of parallelism to explore. We sketched approaches to channel and filter parallelism, which need to be developed further. Reducing inter-node communication overheads for halo exchanges should enable greater spatial partitioning. There remain a host of classic communication optimization techniques, such as overdecomposition, that could be leveraged. Training deep nets on large data is an HPC problem, and tackling it requires exploiting as much parallelism as possible.

\section*{Acknowledgments}
Prepared by LLNL under Contract DE-AC52-07NA27344 (LLNL-CONF-759919). Funding provided by LDRD \#17-SI-003. Some testing/development support work funded by the LLNL Sierra Institutional Center of Excellence. Experiments were performed at the Livermore Computing facility. The authors thank the LBANN and Alkemi teams for their assistance.

\bibliographystyle{IEEEtran}
\bibliography{paper}

\end{document}